\documentclass[journal,twoside,web]{ieeecolor}
\usepackage{jsen}
\usepackage{cite}
\usepackage{amsmath,amssymb,amsfonts,cuted}
\usepackage{algorithm,algorithmic}
\usepackage{graphicx}
\usepackage{textcomp}
\usepackage{wrapfig}
\usepackage{xcolor}
\usepackage[caption=false,font=footnotesize]{subfig}
\usepackage{stfloats}
\usepackage{pgffor}
\usepackage{array}
\usepackage{tabularx}
\usepackage{booktabs,multirow,threeparttable}
\usepackage{makecell}
\usepackage{hyperref}
\usepackage{orcidlink}
\usepackage{caption}
\newcolumntype{C}[1]{>{\centering\arraybackslash}p{#1}}

\def\BibTeX{{\rm B\kern-.05em{\sc i\kern-.025em b}\kern-.08em
    T\kern-.1667em\lower.7ex\hbox{E}\kern-.125emX}}
\definecolor{abstractbg}{rgb}{0.89804,0.94510,0.83137}
\begin{document}

\title{Cuffless Blood Pressure Estimation from Six Wearable Sensor Modalities in Multi-Motion-State Scenarios}

\author{Yiqiao~Chen\textsuperscript{\orcidlink{0009-0003-6503-894X}}, Fazheng~Xu\textsuperscript{\orcidlink{0009-0005-8919-2532}}, Zijian~Huang\textsuperscript{\orcidlink{0000-0001-6849-8827}}, Juchi~He\textsuperscript{\orcidlink{0009-0007-7538-6752}}, and Zhenghui~Feng\textsuperscript{\orcidlink{0000-0003-4230-3053}}
\thanks{Manuscript received Month XX, 2025; revised Month XX, 2025; accepted Month XX, 2025. Date of publication Month XX, 2025; date of current version Month XX, 2025. This work was supported by the Basic Research Fund of the Shenzhen Natural Science Foundation (No. JCYJ20240813104924033). (Corresponding author: Zhenghui Feng.)}
\thanks{Y. Chen, Z. Huang, and Z. Feng are with the Faculty of Frontier Sciences, Harbin Institute of Technology, Shenzhen, Shenzhen 518055, China
(e-mail: cyq2161@gmail.com; 1992oscarhuang@gmail.com; fengzhenghui@hit.edu.cn).}
\thanks{F. Xu is with the University of Melbourne, Australia (e-mail: graycathco3@gmail.com).}
\thanks{J. He is with the University of New South Wales, Australia (stellaris299792458@outlook.com).}
\thanks{Y. Chen, F. Xu, Z. Huang, and J. He contributed equally to this work.}
}

\IEEEtitleabstractindextext{%
\fcolorbox{abstractbg}{abstractbg}{%
\begin{minipage}{\textwidth}%
\begin{wrapfigure}[18]{r}{3in}%
\includegraphics[width=3in]{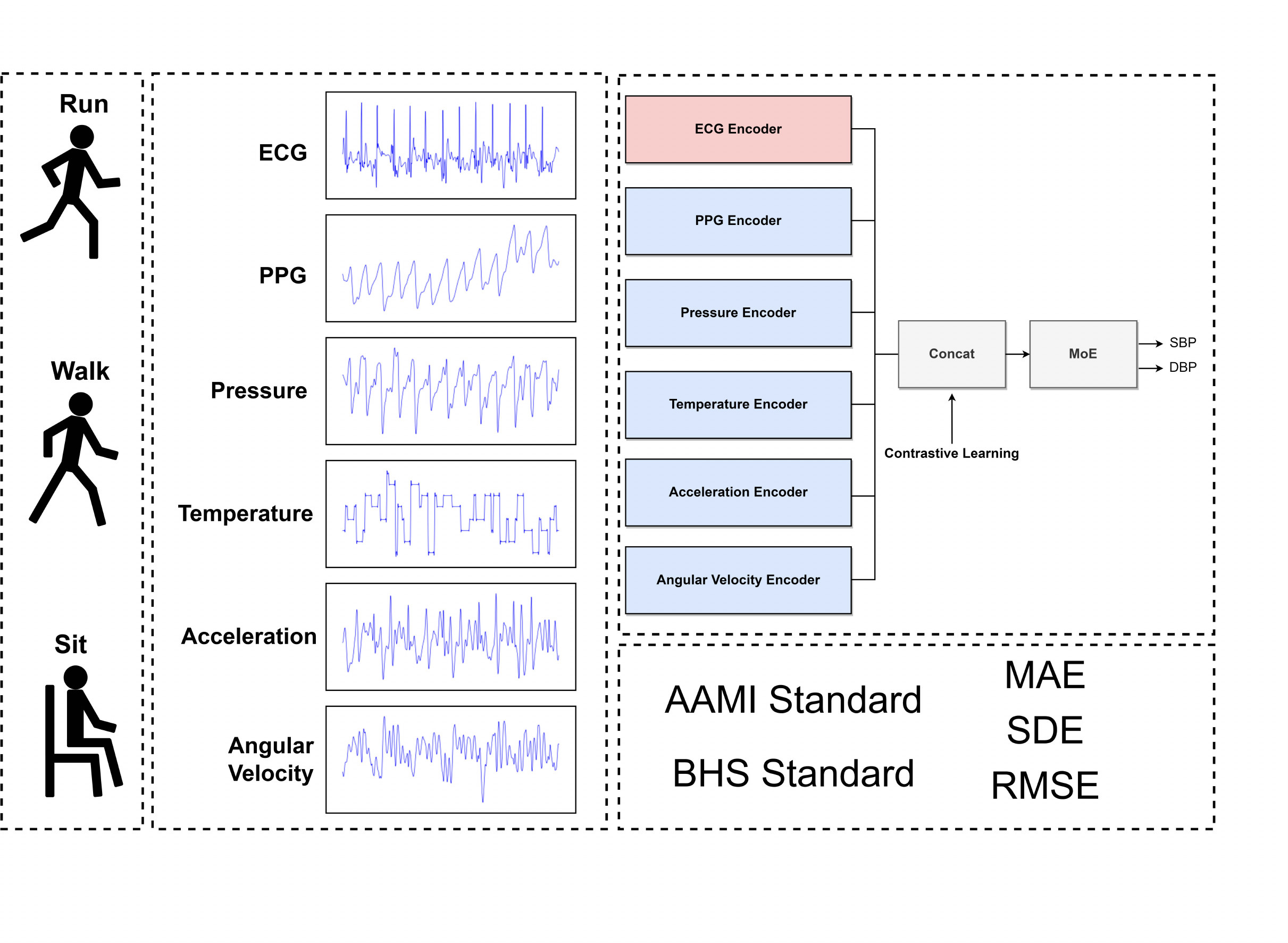}%
\end{wrapfigure}%
\begin{abstract}
Cardiovascular disease (CVD) is a leading cause of morbidity and mortality worldwide, and sustained hypertension is an often silent risk factor, making cuffless continuous blood pressure (BP) monitoring with wearable devices important for early screening and long-term management. Most existing cuffless BP estimation methods use only photoplethysmography (PPG) and electrocardiography (ECG) signals, alone or in combination. These models are typically developed under resting or quasi-static conditions and struggle to maintain robust accuracy in multi-motion-state scenarios. In this study, we propose a six-modal BP estimation framework that jointly leverages ECG, multi-channel PPG, attachment pressure, sensor temperature, and triaxial acceleration and angular velocity. Each modality is processed by a lightweight branch encoder, contrastive learning enforces cross-modal semantic alignment, and a mixture-of-experts (MoE) regression head adaptively maps the fused features to BP across motion states. Comprehensive experiments on the public Pulse Transit Time PPG Dataset, which includes running, walking, and sitting data from 22 subjects, show that the proposed method achieves mean absolute errors (MAE) of 3.60 mmHg for systolic BP (SBP) and 3.01 mmHg for diastolic BP (DBP). From a clinical perspective, it attains Grade A for SBP, DBP, and mean arterial pressure (MAP) according to the British Hypertension Society (BHS) protocol and meets the numerical criteria of the Association for the Advancement of Medical Instrumentation (AAMI) standard for mean error (ME) and standard deviation of error (SDE).
\end{abstract}
\end{minipage}}}

\maketitle

\section{Introduction}

\par Hypertension is one of the most prevalent and important risk factors for cardiovascular disease (CVD)~\cite{dieteren2021prevalence}. According to the World Health Organization (WHO), approximately 1.4 billion adults aged 30--79 years worldwide have hypertension, about 600 million of whom are unaware of their condition, and only around 23\% achieve adequate blood pressure (BP) control~\cite{WHO_hypertension_2025}. Persistently elevated BP accelerates vascular damage and markedly increases the risk of myocardial infarction, stroke, heart failure, renal impairment, and premature death~\cite{Thompson2019}~\cite{marquez2022hypertension}~\cite{shams2025hypertensive}. Although effective lifestyle and pharmacological interventions are available, the global burden of hypertension and related CVD remains unacceptably high. Because hypertension is often asymptomatic~\cite{msd_hypertension}, accurate and repeated BP measurements are crucial for early diagnosis, risk stratification, and long-term management. These considerations underscore the need for BP monitoring technologies that are both reliable and convenient for use in daily life.

\par Upper-arm cuff sphygmomanometers are the current noninvasive standard in clinics and for ambulatory blood pressure monitoring (ABPM), but they only provide intermittent measurements~\cite{muntner2019measurement}~\cite{shin2024evaluation}. Cuff inflation requires the arm to remain still, interrupts daily activities, and repeated nocturnal inflations can disturb sleep and transiently elevate BP, potentially biasing the measurements themselves. Intra-arterial BP monitoring via an arterial catheter offers beat-to-beat accuracy and is widely regarded as the reference method, but it is technically demanding and carries risks such as bleeding, thrombosis and embolism~\cite{owusu2022noninvasive}. As a result, intra-arterial BP monitoring is largely restricted to operating rooms and intensive care units and is unsuitable for routine or home use. These limitations have prompted intensive research into noninvasive, cuffless, and preferably continuous BP monitoring methods.

\par In this context, signal-based cuffless BP estimation has been extensively investigated as a promising alternative to conventional cuff-based sphygmomanometry. Among the various physiological signals explored, photoplethysmography (PPG) and electrocardiography (ECG) have become the most widely used modalities for noninvasive BP estimation~\cite{nie2024review}~\cite{li2022cuffless}. PPG can be conveniently obtained from fingertip probes or wrist-worn devices, whereas ECG is routinely recorded in both clinical practice and wearable monitoring systems. Leveraging these physiological signals, a wide range of computational methods for cuffless BP estimation has been proposed. From a learning-paradigm perspective, early studies typically adopted a hand-crafted feature-engineering pipeline in which morphological, statistical, and frequency-domain features were extracted from PPG and/or ECG and then fed into conventional machine-learning models. Chowdhury et al.~\cite{chowdhury2020estimating} proposed a machine-learning framework that predicts BP from a public single-channel PPG dataset~\cite{liang2018ppgbp} by constructing a feature set of 107 morphological, frequency-domain, statistical, and demographic descriptors. Using Gaussian process regression (GPR) together with ReliefF feature selection, their best model achieved root-mean-square errors (RMSEs) of 6.74~mmHg and 3.59~mmHg for systolic and diastolic BP, respectively. Aguet et al.~\cite{aguet2023blood} proposed a PPG-based BP-monitoring approach during anesthesia induction that uses pulse-wave morphology features and patient demographics, combined with machine-learning regressors such as least absolute shrinkage and selection operator (Lasso), support vector regression (SVR), and GPR. On a dataset of 40 surgical patients with invasive arterial BP as reference, their calibrated Lasso model achieved mean error (ME) $\pm$ standard deviation of error (SDE) of $-0.87 \pm 10.77$~mmHg and $-1.31 \pm 7.62$~mmHg for systolic and diastolic BP, respectively.

\par However, the performance of such pipelines strongly depends on the choice of hand-crafted features and prior physiological assumptions, which may limit their ability to capture complex nonlinear dynamics and cross-beat variability in PPG and ECG signals. In recent years, end-to-end deep learning approaches have therefore been explored to automatically learn task-relevant representations directly from raw or minimally processed signals, offering greater flexibility for modeling temporal dependencies and nonlinear relationships in cuffless BP estimation. Rong and Li~\cite{rong2021multi} proposed a multi-type feature fusion neural network that predicts BP from PPG by combining morphological, spectral, and temporal features through parallel convolutional neural network (CNN) and bidirectional long short-term memory (BLSTM) branches. On the UCI cuffless BP dataset, their fusion model achieved mean error (ME) $\pm$ standard deviation (SD) of $-1.13 \pm 7.25$~mmHg and $0.14 \pm 4.48$~mmHg for systolic and diastolic BP, respectively. Kim et al.~\cite{kim2022deepcnap} proposed DeepCNAP, a deep learning model that combines deep convolution and self-attention to estimate continuous arterial BP waveforms from PPG signals. On a UCI cuffless BP dataset derived from MIMIC-II with 942 subjects and 10-fold cross-validation, DeepCNAP achieved mean absolute error (MAE) $\pm$ SD of $3.40 \pm 4.36$~mmHg and $1.75 \pm 2.25$~mmHg for systolic and diastolic BP, respectively. Shaikh and Forouzanfar~\cite{shaikh2024dual} proposed a dual-stream convolutional neural network--long short-term memory (CNN-LSTM) network for cuffless BP estimation using synchronized PPG and ECG signals; on the PulseDB dataset with more than 3000 subjects, their model achieved MAEs of 5.16~mmHg and 3.24~mmHg for systolic and diastolic BP, respectively. Kamanditya et al.~\cite{kamanditya2024continuous} proposed a Conv1D--LSTM-based continuous BP prediction system that uses short segments of PPG and ECG signals together with R-to-R interval durations, trained on the MIMIC-III dataset; their best configuration achieved MAE $\pm$ SD of $5.306 \pm 7.248$~mmHg and $3.296 \pm 4.764$~mmHg for systolic and diastolic BP, respectively. Suhas BN et al.~\cite{suhas2024end} proposed an end-to-end transformer-based model for personalized cuffless BP monitoring from raw ECG and PPG signals, trained with a weighted contrastive loss in addition to MAE, and their subject-specific models on PulseDB achieved average MAEs of 1.08~mmHg and 0.68~mmHg for systolic and diastolic BP, respectively. Kumar et al.~\cite{kumar2024swin} proposed a Swin Transformer-based method that estimates BP from bispectrum-derived higher-order spectral features of ECG and PPG signals; using a UCI cuffless BP dataset derived from MIMIC-II, they reported MAE $\pm$ SD of $5.72 \pm 5.22$~mmHg and $2.64 \pm 3.45$~mmHg for systolic and diastolic BP, respectively. Tian et al.~\cite{tian2025paralleled} proposed a paralleled CNN-Transformer network (PCTN) for PPG-based cuffless BP estimation, and on the MIMIC database it achieved mean absolute errors of 2.36~mmHg, 2.03~mmHg, and 4.44~mmHg for diastolic, mean arterial, and systolic BP, respectively.

\par Existing studies on cuffless BP estimation differ markedly in their network architectures and learning paradigms, but they also exhibit clear patterns in terms of which physiological signals are used as inputs. From the perspective of sensing configuration, prior work can be broadly grouped by the number and type of modalities involved. A first line of studies adopts a single-modality configuration, most commonly relying on PPG alone and occasionally on ECG alone~\cite{chowdhury2020estimating}~\cite{kim2022deepcnap}~\cite{aguet2023blood}~\cite{rong2021multi}~\cite{tazarv2021deep}~\cite{haddad2021continuous}~\cite{cao2021crisp}. To enrich the available information, other studies augment PPG with an additional modality, most frequently ECG, thereby forming a two-signal configuration~\cite{kumar2024swin}~\cite{yen2022estimation}~\cite{suhas2024end}~\cite{shaikh2024dual}~\cite{kamanditya2024continuous}. Overall, most existing works employ only one or two physiological modalities and are primarily developed and evaluated under resting or quasi-static conditions~\cite{hove2024prototype}.

\par In realistic daily-life use, however, wearable devices must operate during activities such as sitting, walking, and running, where PPG and ECG signals change substantially across motion states and with variations in sensor–skin contact, which markedly degrade signal quality and hinder reliable physiological parameter estimation~\cite{al2023adaptive}. At the same time, modern wearable platforms typically also record attachment pressure, sensor temperature, and inertial measurement unit (IMU) signals, including triaxial acceleration and angular velocity. Rather than being treated as secondary channels, these additional signals can provide complementary contextual information about contact stability and body motion that is helpful for interpreting wearable cardiovascular waveforms in multi-motion-state scenarios. However, in many systems described as “multimodal”, such sensing channels are either not integrated into the BP estimation model at all or are only used in a limited auxiliary manner. Consequently, the potential of richer multimodal sensing for improving the robustness of cuffless BP estimation across multi-motion-state scenarios remains largely underexplored, highlighting the need for modeling frameworks that explicitly integrate multiple wearable sensor modalities and are systematically validated under diverse motion states.

\par Motivated by the limited use of richer wearable sensing and the lack of systematic validation in multi-motion-state scenarios, we propose a six-modal framework for cuffless BP estimation based on the Pulse Transit Time PPG Dataset~\cite{mehrgardt2022pulse}. We first construct a multi-motion-state dataset by segmenting the synchronized recordings from running, walking, and sitting into fixed-length windows and applying a widely used hybrid denoising method that combines variational mode decomposition and wavelet transform (VMD–WT)~\cite{qin2023novel}, thereby obtaining six time-aligned sensor modalities (ECG, multi-channel PPG, attachment pressure, sensor temperature, and triaxial acceleration and angular velocity) in a unified representation space. On top of this preprocessed data, we design a six-branch encoder that learns modality-specific features, followed by a contrastive-learning-based semantic alignment fusion module and a Mixture-of-Experts (MoE) regression head that jointly estimate systolic and diastolic BP across motion states. 

\par The main contributions of this work are summarized as follows:
\begin{enumerate}
\item We investigate, to the best of our knowledge, one of the first cuffless BP estimation frameworks that simultaneously exploits six synchronized wearable sensor modalities—ECG, multi-channel PPG, attachment pressure, sensor temperature, and inertial signals (triaxial acceleration and angular velocity)—specifically targeting multi-motion-state scenarios.
\item We construct a multi-motion-state dataset from the Pulse Transit Time PPG Dataset by segmenting running, walking, and sitting recordings and applying a unified preprocessing pipeline (including resampling and noise suppression based on a Variational Mode Decomposition–Wavelet Transform (VMD–WT) scheme), yielding time-aligned six-modal data suitable for deep multimodal learning and subsequent research on motion-robust cuffless BP estimation.
\item We propose a six-branch multimodal architecture that combines modality-specific encoders, a contrastive-learning-based semantic alignment fusion module, and a Mixture-of-Experts (MoE) regression head, enabling adaptive integration of heterogeneous modalities for BP estimation across motion states.
\item Through extensive experiments under running, walking, and sitting conditions, we show that the proposed framework consistently outperforms single- and dual-modality baselines on the Pulse Transit Time PPG Dataset. The model achieves MAEs of 3.60~mmHg and 3.01~mmHg for systolic and diastolic BP, respectively, attains Grade~A for systolic, diastolic, and mean arterial pressure according to the British Hypertension Society (BHS) protocol, and meets the numerical criteria of the Association for the Advancement of Medical Instrumentation (AAMI) standard in terms of mean error and standard deviation of error.
\end{enumerate}

\par The remainder of this paper is organized as follows. Section~\ref{sec:met} describes the Pulse Transit Time PPG Dataset, the multi-motion-state preprocessing pipeline, and the proposed six-modal cuffless BP estimation framework, as well as the experimental setup. Section~\ref{sec:exp} reports the experimental results across different motion states and provides a detailed discussion of the findings and limitations. Finally, Section~\ref{sec:con} concludes the paper and outlines directions for future research.

\section{Methodology}\label{sec:met}
\subsection{Dataset}
\begin{table*}[htbp]
	\centering
	\caption{Summary of the 18 time-synchronized sensor channels used in this work, including ECG, multi-wavelength PPG from two finger segments, load-cell attachment pressure, sensor temperature, and inertial measurements.}
    \label{tab:data_channels}
	\small
	\setlength{\tabcolsep}{4pt}
	\renewcommand{\arraystretch}{1.2}
	\begin{tabular}{@{}llr@{}}
		\toprule
		\textbf{Name} & \textbf{Description} & \textbf{Sampling Rate} \\
		\midrule
		\multicolumn{3}{@{}l}{\textbf{Electrocardiography (ECG)}} \\
		ecg      & Single-channel ECG signal (three-lead configuration) & 500 Hz \\
		\midrule
		\multicolumn{3}{@{}l}{\textbf{Photoplethysmography (PPG)}} \\
		pleth\_1 & MAX30101 red wavelength PPG from the distal phalanx of the left index finger palmar side & 500 Hz \\
		pleth\_2 & MAX30101 infrared wavelength PPG from the distal phalanx of the left index finger palmar side & 500 Hz \\
		pleth\_3 & MAX30101 green wavelength PPG from the distal phalanx of the left index finger palmar side & 500 Hz \\
		pleth\_4 & MAX30101 red wavelength PPG from the proximal phalanx of the left index finger palmar side & 500 Hz \\
		pleth\_5 & MAX30101 infrared wavelength PPG from the proximal phalanx  of the left index finger palmar side & 500 Hz \\
		pleth\_6 & MAX30101 green wavelength PPG from the proximal phalanx of the left index finger palmar side & 500 Hz \\
		\midrule
		\multicolumn{3}{@{}l}{\textbf{Sensor Physical Parameters}} \\
		lc\_1    & TAL221 load cell proximal phalanx (first segment) PPG sensor attachment pressure & 80 Hz \\
		lc\_2    & TAL221 load cell (base segment) PPG sensor attachment pressure & 80 Hz \\
		temp\_1  & Distal phalanx (first segment) PPG sensor temperature & 10 Hz \\
		temp\_2  & Proximal phalanx (base segment) PPG sensor temperature & 10 Hz \\
		temp\_3  & InvenSenseMPU-9250 IMU temperature & 500 Hz \\
		\midrule
		\multicolumn{3}{@{}l}{\textbf{Inertial Measurement Unit}} \\
		a\_x & InvenSenseMPU-9250 IMU acceleration in x-direction & 500 Hz \\
		a\_y & InvenSenseMPU-9250 IMU acceleration in y-direction & 500 Hz \\
		a\_z & InvenSenseMPU-9250 IMU acceleration in z-direction & 500 Hz \\
		g\_x & InvenSenseMPU-9250 IMU angular velocity around x-axis & 500 Hz \\
		g\_y & InvenSenseMPU-9250 IMU angular velocity around y-axis & 500 Hz \\
		g\_z & InvenSenseMPU-9250 IMU angular velocity around z-axis & 500 Hz \\
		\bottomrule
	\end{tabular}
\end{table*}
\par This study uses the Pulse Transit Time PPG Dataset (version 1.1.0). The dataset comprises 66 recording files containing 19 time-synchronized sensor channels and numerical reference values, acquired from 22 healthy subjects during three motion states: running, sitting, and walking. It was originally designed to support research on short-distance pulse transit time recognition, cuffless blood pressure sensing, and cardiovascular modeling~\cite{mehrgardt2022pulse}. 

\par In this work, we use 18 wearable sensor channels as model inputs, including one ECG channel sampled at 500~Hz with a three-lead configuration, six PPG channels from two finger segments and three wavelengths, two load-cell channels measuring sensor attachment pressure, three sensor temperature channels, and six inertial measurement unit (IMU) channels capturing triaxial acceleration and angular velocity. A summary of these input channels and their sampling rates is given in Table~\ref{tab:data_channels}.

\subsection{Data Processing Pipeline}
\begin{figure}[htbp]
    \centering
    \includegraphics[width=0.45\textwidth]{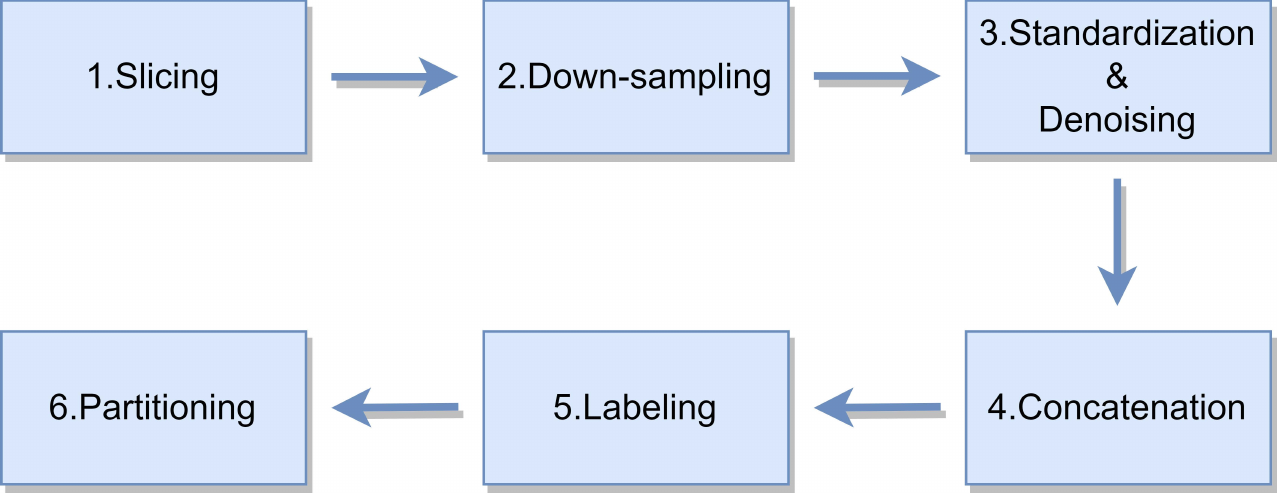}
    \caption{Preprocessing pipeline used to construct the six-modal multi-motion-state dataset from the Pulse Transit Time PPG recordings, including slicing, down-sampling, standardization and VMD–WT denoising, concatenation, labeling, and partitioning.}
    \label{fig:preprocess_pipeline}
\end{figure}

\begin{figure}[htbp]
    \centering
    \includegraphics[width=0.45\textwidth]{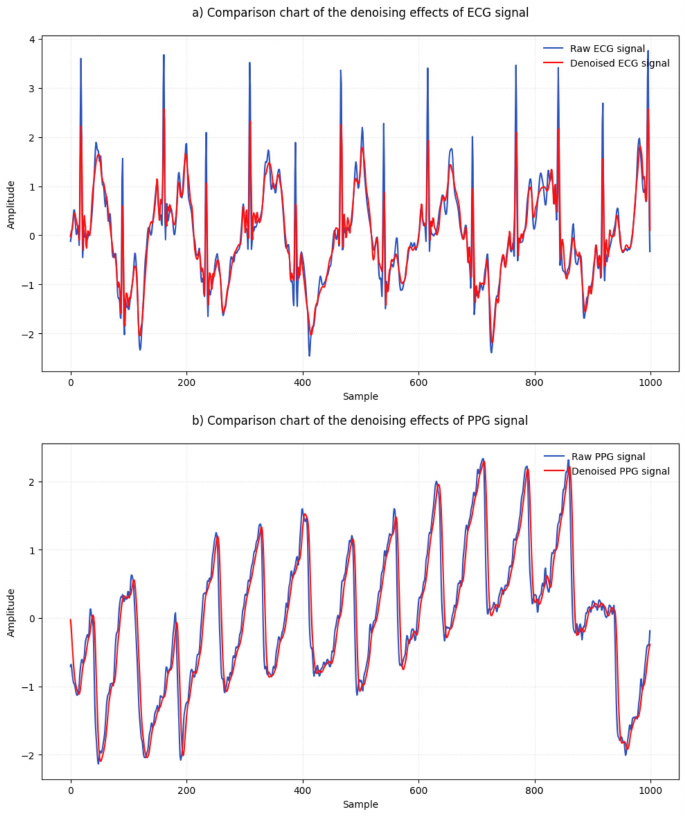}
    \caption{Illustrative example of the denoising pipeline applied to ECG and PPG signals, comparing raw waveforms with the corresponding filtered and denoised signals.}
    \label{fig:denoise_example}
\end{figure}

\par To obtain learning-ready, time-aligned segments from the raw Pulse Transit Time PPG recordings, we implement a six-stage preprocessing pipeline, summarized in Fig.~\ref{fig:preprocess_pipeline}. Specifically, for each subject and motion state (running, walking, and sitting), we first retrieved the corresponding recording files, removed the auxiliary \texttt{peaks} sequence, and segmented the remaining 18 sensor channels into non-overlapping windows of 5000 samples. For the ECG channel, which is sampled at 500~Hz, each 5000-sample window spans 10~s and thus typically covers at least 10 cardiac cycles, providing sufficient temporal context for BP estimation. To obtain a unified temporal resolution and reduce computational load, we treated 500~Hz as the common highest effective sampling rate across channels (ECG, PPG, and IMU were recorded at 500~Hz, whereas the attachment-pressure and PPG sensor-temperature channels were up-sampled to 500~Hz in the released dataset) and then applied a fourth-order Butterworth low-pass filter with a 50~Hz cutoff followed by decimation by a factor of five, yielding 100~Hz segments with 1000 samples per channel.

\par To mitigate the impact of amplitude differences across channels and segments on model training, we then standardized each signal on a per-segment basis by subtracting the segment-wise mean and dividing by the segment-wise standard deviation. Because biomedical waveforms are inevitably affected by motion artefacts, baseline drift, and environmental noise, we applied dedicated denoising to the ECG and PPG channels. For ECG, we first used a fourth-order Butterworth low-pass filter with a 40~Hz cutoff frequency to further suppress high-frequency environmental and instrumental noise, and then adopted a hybrid VMD--WT denoising framework that has been widely used for nonstationary biomedical signals~\cite{dragomiretskiy2013variational,donoho1994ideal,qin2023novel}. Concretely, each standardized ECG segment was decomposed into $K=6$ intrinsic mode functions (IMFs) using Variational Mode Decomposition (VMD)~\cite{dragomiretskiy2013variational}; each IMF was subsequently processed by wavelet shrinkage using a Daubechies-4 (db4) wavelet basis with a decomposition level of 4~\cite{donoho1994ideal}, and the denoised IMFs were recombined to obtain a cleaned ECG segment. For the multi-channel PPG signals, each channel was individually processed using a second-order Butterworth low-pass filter with a cutoff frequency of 7 Hz. No additional denoising was applied to the remaining signals. The denoising effects of ECG and PPG signals are illustrated in Fig.~\ref{fig:denoise_example}.

\par Subsequently, for each preprocessed segment we grouped channels belonging to the same sensor type to construct six modality-specific inputs:
(i) a single-channel ECG signal $x^{(e)} \in \mathbb{R}^{1 \times 1000}$,
(ii) a six-channel PPG signal $x^{(p)} \in \mathbb{R}^{6 \times 1000}$,
(iii) a two-channel attachment-pressure signal $x^{(l)} \in \mathbb{R}^{2 \times 1000}$,
(iv) a three-channel temperature signal $x^{(t)} \in \mathbb{R}^{3 \times 1000}$,
(v) a three-channel acceleration signal $x^{(a)} \in \mathbb{R}^{3 \times 1000}$, and
(vi) a three-channel angular-velocity signal $x^{(g)} \in \mathbb{R}^{3 \times 1000}$.
Each sample thus consists of six temporally aligned modalities with a shared segment length of 1000 time steps.

\par Finally, because the Pulse Transit Time PPG Dataset provides only two cuff-based BP measurements (at the beginning and at the end of each recording), all segments extracted from a given subject–state recording (e.g., s1\_run) were assigned the corresponding bp\_sys\_end and bp\_dia\_end values from the annotation file as approximate ground-truth labels for systolic and diastolic BP, respectively. The resulting segment-level samples from all subjects were then randomly partitioned, on a segment-wise basis, into training, validation, and test sets with a ratio of 7:1:2 for subsequent model development and evaluation.

\subsection{Model Architecture}
\begin{figure*}[htbp]
    \centering
    \includegraphics[width=1\textwidth]{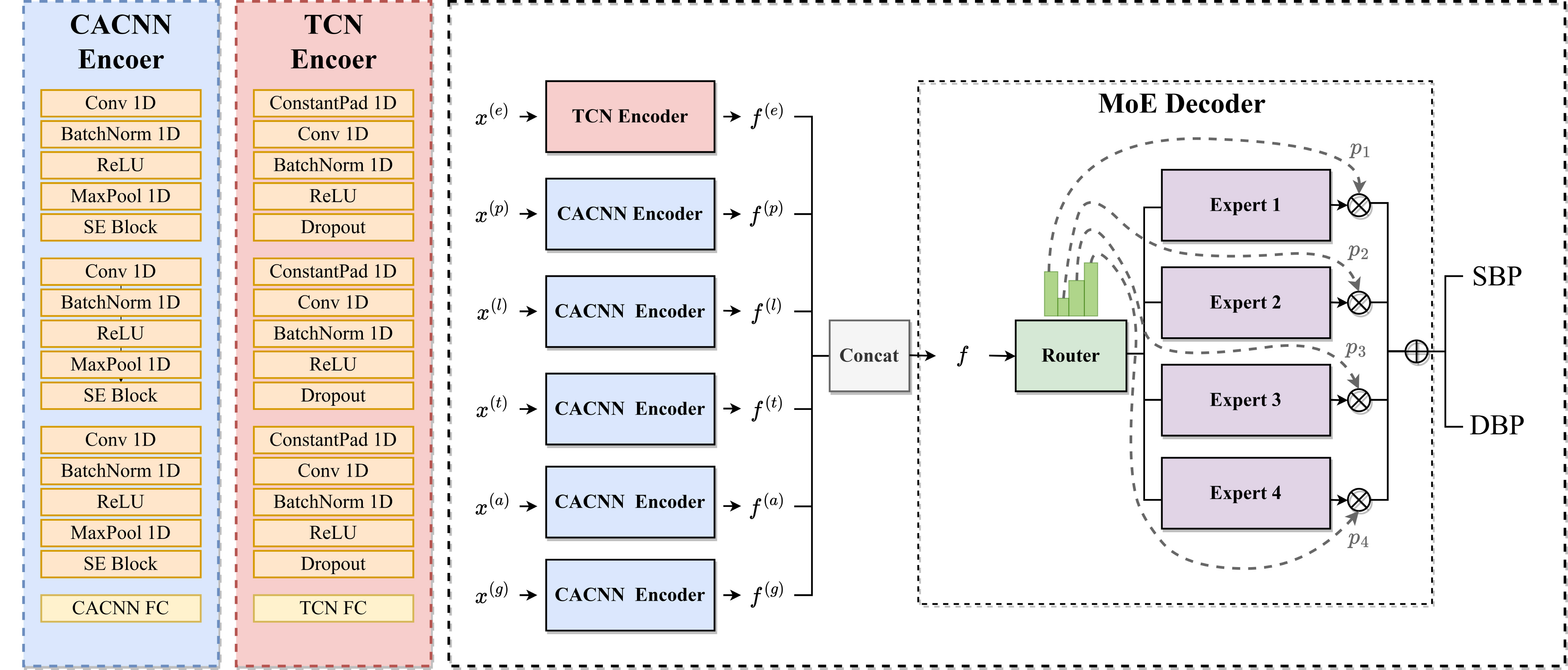}
    \caption{Overall architecture of the proposed six-modal model, including six modality-specific encoders, a contrastive-learning-based cross-modal semantic alignment module, and a Mixture-of-Experts (MoE) regression head for joint systolic and diastolic blood pressure estimation.}
    \label{fig:ctam_architecture}
\end{figure*}

\begin{figure}[htbp]
    \centering
    \includegraphics[width=0.45\textwidth]{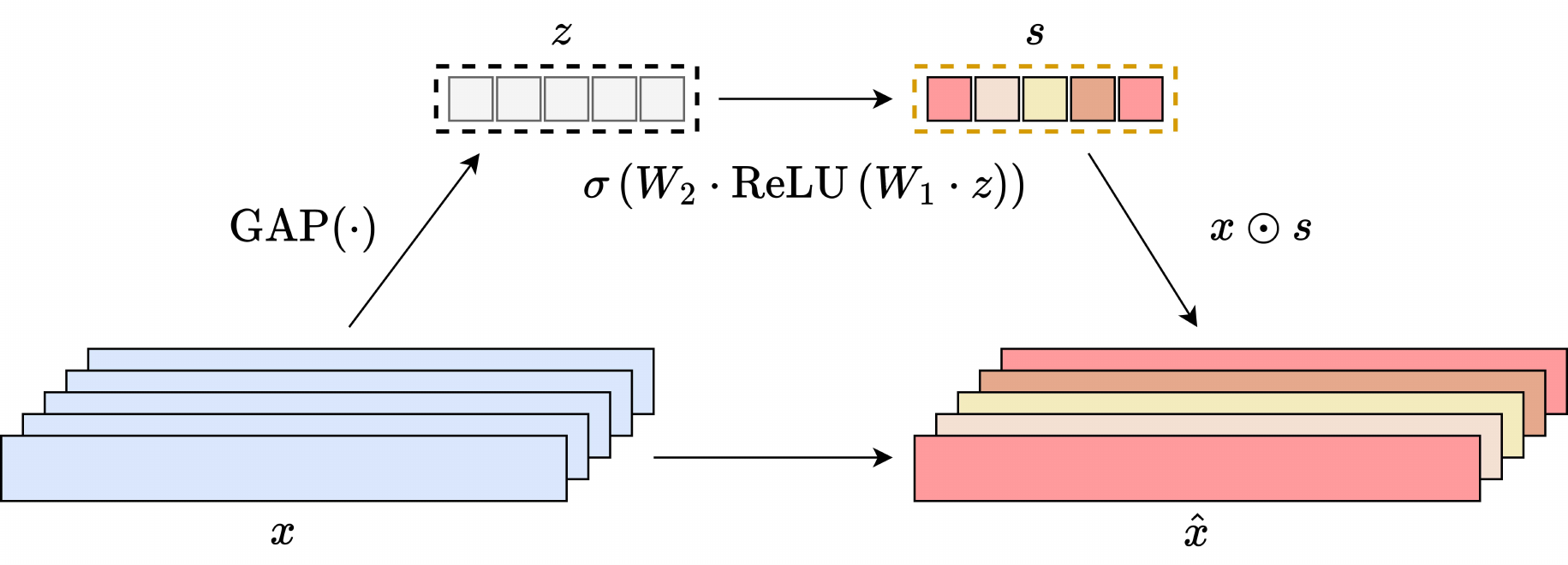}
    \caption{Structure of the one-dimensional squeeze-and-excitation (SE) block used in the CACNN encoder to implement channel-wise attention for multi-channel sensor inputs.}
    \label{fig:se_block}
\end{figure}

\begin{table*}[htbp]
	\centering
	\caption{Architecture of the temporal convolutional network (TCN) encoder branch used to extract features from the single-lead ECG signal.}
    \label{tab:tcn_encoder}
	\begin{tabularx}{\textwidth}{>{\bfseries}l l X}
		\toprule
		Module      & Layer           & Parameters \\
		\midrule
		& ConstantPad1D   & Left = 2, Right = 0 \\
		& Conv1D          & Kernel Size = 3, In Channels = 1, Out Channels = 128, Stride = 1, Dilation = 1 \\
		Level1 TCN  & BatchNorm1D     & Num Features = 128 \\
		& ReLU            & --- \\
		& Dropout         & p = 0.2 \\
		\midrule
		& ConstantPad1D   & Left = 4, Right = 0 \\
		& Conv1D          & Kernel Size = 3, In Channels = 128, Out Channels = 64, Stride = 1, Dilation = 2 \\
		Level2 TCN  & BatchNorm1D     & Num Features = 64 \\
		& ReLU            & --- \\
		& Dropout         & p = 0.2 \\
		\midrule
		& ConstantPad1D   & Left = 8, Right = 0 \\
		& Conv1D          & Kernel Size = 3, In Channels = 64, Out Channels = 9, Stride = 1, Dilation = 4 \\
		Level3 TCN  & BatchNorm1D     & Num Features = 9 \\
		& ReLU            & --- \\
		& Dropout         & p = 0.2 \\
		\midrule
		& Flatten         & --- \\
		& Linear          & In Features = 9000, Out Features = 4096 \\
		& ReLU            & --- \\
		& Dropout         & p = 0.2 \\
		& Linear          & In Features = 4096, Out Features = 2048 \\
		TCN FC      & ReLU            & --- \\
		& Dropout         & p = 0.2 \\
		& Linear          & In Features = 2048, Out Features = 512 \\
		& ReLU            & --- \\
		& Dropout         & p = 0.2 \\
		& Linear          & In Features = 512, Out Features = 128 \\
		\bottomrule
	\end{tabularx}
\end{table*}

\begin{table*}[htbp]
	\centering
	\caption{Architecture of the channel-attention convolutional neural network (CACNN) encoder for multi-channel PPG and physical sensor modalities (load-cell pressure, temperature, acceleration, and angular velocity).}
    \label{tab:cacnn_encoder}
	\begin{tabularx}{\textwidth}{>{\bfseries}l l X}
		\toprule
		Module      & Layer           & Parameters \\
		\midrule
		& Conv1D   & Kernel Size = 18, In Channels = $x^{(p)}:6,x^{(l)}:2,x^{(t)}:3,x^{(a)}:3,x^{(g)}:3$, Out Channels = 27, Stride = 1, Dilation = 1, Padding = 0 \\
		& BatchNorm1D          & Num Features = 27 \\
		Level1 CACNN  & ReLU     & --- \\
		& MaxPool1D            & Kernel Size = 3, Stride = 3 \\
		& SE Block         & Reduction Ratio = 9 \\
		\midrule
		& Conv1D   & Kernel Size = 9, In Channels = 27, Out Channels = 27, Stride = 1, Dilation = 1, Padding = 0 \\
		& BatchNorm1D          & Num Features = 27 \\
		Level2 CACNN  & ReLU     & --- \\
		& MaxPool1D            & Kernel Size = 3, Stride = 3 \\
		& SE Block         & Reduction Ratio = 9 \\
		\midrule
		& Conv1D   & Kernel Size = 7, In Channels = 27, Out Channels = 27, Stride = 1, Dilation = 1, Padding = 0 \\
		& BatchNorm1D          & Num Features = 27 \\
		Level3 CACNN  & ReLU     & --- \\
		& MaxPool1D            & Kernel Size = 3, Stride = 3 \\
		& SE Block         & Reduction Ratio = 9 \\
		\midrule
		& Flatten         & --- \\
		& Linear          & In Features = 891, Out Features = 512 \\
		& ReLU            & --- \\
		& Dropout         & p = 0.3 \\
		& Linear          & In Features = 512, Out Features = 512 \\
		CACNN FC      & ReLU            & --- \\
		& Dropout         & p = 0.3 \\
		& Linear          & In Features = 512, Out Features = 256 \\
		& ReLU            & --- \\
		& Dropout         & p = 0.3 \\
		& Linear          & In Features = 256, Out Features = 128 \\
		\bottomrule
	\end{tabularx}
\end{table*}

\begin{table}[htbp]
	\centering
	\caption{Layer configuration of the multilayer perceptron (MLP) expert network in the Mixture-of-Experts (MoE) regression head for joint SBP/DBP prediction.}
    \label{tab:mlp_expert}
	\begin{tabularx}{0.5\textwidth}{>{\bfseries}l X}
		\toprule
		Layer & Parameters \\
		\midrule
		Linear & In Features = 768, Out Features = 512\\
		ReLU & --- \\
		BatchNorm & Num Features = 512\\
		Linear & In Features = 512, Out Features = 512\\
		ReLU & --- \\
		BatchNorm & Num Features = 512\\
		Linear & In Features = 512, Out Features = 256\\
		ReLU & --- \\
		BatchNorm & Num Features = 256\\
		Linear & In Features = 256, Out Features = 2\\
		\bottomrule
	\end{tabularx}
\end{table}

\par As shown in Fig.~\ref{fig:ctam_architecture}, the proposed model consists of six modality-specific encoders, a contrastive-learning-based cross-modal semantic alignment module, and a Mixture-of-Experts (MoE) regression head for joint SBP/DBP estimation. To capture temporal information in the single-channel ECG, including waveform morphology and timing intervals of P waves, QRS complexes, and T waves, we employ a temporal convolutional network (TCN) encoder based on causal and dilated convolutions~\cite{bai2018empirical}. To keep the model depth and computational cost moderate, we omit the residual connections used in the original TCN formulation and adopt a compact three-level design. The detailed layer configuration of the TCN encoder is summarized in Table~\ref{tab:tcn_encoder}.

\par For the remaining five modalities (multi-channel PPG, attachment pressure, temperature, acceleration, and angular velocity), we use a channel-attention convolutional neural network (CACNN) encoder to extract waveform features and model inter-channel relationships. Each CACNN block incorporates a squeeze-and-excitation (SE) channel-attention module~\cite{hu2018squeeze}, which applies global average pooling followed by a lightweight MLP with sigmoid gating to generate channel-wise attention weights that rescale the feature maps, adaptively emphasizing informative channels while suppressing less relevant ones (cf. Fig.~\ref{fig:se_block} / corresponding SE illustration). The overall architecture of the CACNN encoder is listed in Table~\ref{tab:cacnn_encoder}.

\par Building on the TCN and CACNN encoders described above, the proposed model processes each modality through its corresponding encoder to obtain modality-specific representations. For a given sample, feature extraction is performed as
\begin{align}
    f^{(e)} &= \text{TCNEncoder}\bigl(x^{(e)}\bigr) \in \mathbb{R}^{128}, \\
    f^{(p)} &= \text{CACNNEncoder}\bigl(x^{(p)}\bigr) \in \mathbb{R}^{128}, \\
    f^{(l)} &= \text{CACNNEncoder}\bigl(x^{(l)}\bigr) \in \mathbb{R}^{128}, \\
    f^{(t)} &= \text{CACNNEncoder}\bigl(x^{(t)}\bigr) \in \mathbb{R}^{128}, \\
    f^{(a)} &= \text{CACNNEncoder}\bigl(x^{(a)}\bigr) \in \mathbb{R}^{128}, \\
    f^{(g)} &= \text{CACNNEncoder}\bigl(x^{(g)}\bigr) \in \mathbb{R}^{128},
\end{align}
where $f^{(e)}, f^{(p)}, f^{(l)}, f^{(t)}, f^{(a)}, f^{(g)}$ denote the 128-dimensional feature vectors encoded from the six modality-specific inputs $x^{(e)}, x^{(p)}, x^{(l)}, x^{(t)}, x^{(a)}, x^{(g)}$. These vectors are then concatenated to form a joint representation
\begin{equation}\label{concat}
    f =
    \begin{bmatrix}
        f^{(e)} \\
        f^{(p)} \\
        f^{(l)} \\
        f^{(t)} \\
        f^{(a)} \\
        f^{(g)}
    \end{bmatrix}
    \in \mathbb{R}^{768}.
\end{equation}
Finally, the concatenated feature vector $f \in \mathbb{R}^{768}$ is fed into a Mixture-of-Experts (MoE) decoder to predict systolic and diastolic BP:
\begin{equation}
    \begin{bmatrix}
        \mathrm{SBP} \\
        \mathrm{DBP}
    \end{bmatrix}
    = \text{MoEDecoder}(f)
    = \sum_{i=1}^{E} g_i(f)\, E_i(f),
\end{equation}
where $E$ is the number of experts, $g_i(f)$ is the gating score of the $i$-th expert produced by a softmax gating network, and $E_i(\cdot)$ denotes the $i$-th expert. Each expert $E_i$ is implemented as a four-layer MLP with hidden dimensions 512, 512, and 256 and an output dimension of 2, as summarized in Table~\ref{tab:mlp_expert}. Compared with using a single monolithic regressor on the fused representation $f$, the MoE head is introduced to more flexibly model the heterogeneous relationships between the six-modal features and blood pressure across different motion states. In practice, the concatenated feature vector $f$ in (\ref{concat}) may correspond to substantially different regimes of motion, sensor--skin contact, and subject-specific physiological patterns, which makes a single shared mapping from $f$ to $(\mathrm{SBP}, \mathrm{DBP})$ suboptimal. By contrast, in the MoE design each expert $E_i$ is encouraged to specialize in a particular region of the feature space, while the gating network $g_i(f)$ adaptively produces sample-dependent mixture weights based on the current fused features. The final prediction $\sum_{i=1}^{E} g_i(f)\,E_i(f)$ can thus be interpreted as a data-dependent ensemble of specialized regressors, rather than the output of a single global model. This increases the functional capacity of the regression module without simply stacking more layers, and allows the network to learn more flexible, piecewise relationships between the six-modal representation and blood pressure in complex multi-motion-state scenarios.

\subsection{Learning Strategy}
To train the proposed model, we combine a supervised regression loss with a six-modal contrastive loss. First, we use the standard mean squared error (MSE) between the predicted and reference BP values. Let $\mathbf{y}_i \in \mathbb{R}^2$ and $\hat{\mathbf{y}}_i \in \mathbb{R}^2$ denote the ground-truth and predicted $(\mathrm{SBP}, \mathrm{DBP})$ for the $i$-th sample in a mini-batch of size $B$. The MSE loss is given by
\begin{equation}
    \mathcal{L}_{\text{MSE}}
    = \frac{1}{B} \sum_{i=1}^{B} 
      \bigl\|\mathbf{y}_i - \hat{\mathbf{y}}_i\bigr\|_2^2.
\end{equation}

\par In addition, because the six modality-specific feature vectors of a given sample originate from the same underlying cardiovascular and motion state and are therefore expected to share a common high-level semantics, we introduce a contrastive loss over all unordered modality pairs to encourage semantic alignment in the embedding space and to mitigate modality-specific noise and motion artefacts. This loss pulls together embeddings from different modalities of the same sample while pushing apart embeddings originating from different samples, thereby promoting a more coherent shared representation for subsequent multimodal fusion and blood pressure regression. For a given sample $i$, let $\{f_i^{(m)}\}_{m=1}^{6}$ denote the six modality embeddings. For each unordered pair of modalities $(p,q) \in \binom{6}{2}$, we treat $(f_i^{(p)}, f_i^{(q)})$ as a positive pair and form $K$ negative pairs by randomly pairing $f_i^{(p)}$ with embeddings $f_j^{(q)}$ from other samples $j \neq i$ in the batch. Cosine similarity between two embeddings $f^p$ and $f^q$ is defined as
\begin{equation}
    d = \operatorname{sim}(f^p, f^q)
      = \frac{f^p \cdot f^q}{\|f^p\| \, \|f^q\|}.
\end{equation}
Denoting by $d_{\text{pos}}^{(i,p,q)}$ the similarity of the positive pair for sample $i$ and modality pair $(p,q)$, and by $d_{\text{neg},j}^{(i,p,q)}$ the similarities of the $K$ negative pairs, we adopt an InfoNCE-style objective~\cite{he2020momentum}:
\begin{equation}
    \mathcal{L}_{p,q}
    = -\frac{1}{B} \sum_{i=1}^{B} 
    \log 
    \frac{
        \exp\!\left( \dfrac{d_{\text{pos}}^{(i,p,q)}}{\tau} \right)
    }{
        \exp\!\left( \dfrac{d_{\text{pos}}^{(i,p,q)}}{\tau} \right)
        + 
        \sum_{j=1}^{K} \exp\!\left( \dfrac{d_{\text{neg},j}^{(i,p,q)}}{\tau} \right)
    },
\end{equation}
where $\tau > 0$ is a temperature parameter. The six-modal contrastive loss is obtained by averaging over all $\binom{6}{2}=15$ modality pairs:
\begin{equation}
    \mathcal{L}_{\text{Contrastive}}
    = \frac{1}{15} 
      \sum_{(p,q) \in \binom{6}{2}} \mathcal{L}_{p,q}.
\end{equation}

\par Combining the MSE loss and the contrastive loss, the overall training objective is
\begin{equation}
    \mathcal{L}_{\text{total}}
    = \mathcal{L}_{\text{MSE}} 
      + \lambda \, \mathcal{L}_{\text{Contrastive}},
\end{equation}
where $\lambda > 0$ controls the relative weight of the contrastive term.

\subsection{Experimental Setup}
\par Experiments were conducted using two AMD EPYC 7B12 CPUs (64 cores, 128 threads, 2.25~GHz base frequency) and eight NVIDIA RTX~4090 GPUs (24~GB memory per GPU, 16{,}384 CUDA cores). All models were implemented in PyTorch. During training, we used a mini-batch size of 24 and optimized the network parameters with the Adam~\cite{kingma2014adam} optimizer with an initial learning rate of $3\times10^{-4}$. The proposed model was trained for 100~epochs. The training objective combined the mean squared error (MSE) regression loss and the six-modal contrastive InfoNCE loss described in the previous subsection, with a weighting factor $\lambda = 0.3$ on the contrastive loss, a temperature parameter $\tau = 0.5$, and $K = 5$ negative pairs sampled per positive pair.

\par To establish strong baselines for cuffless BP estimation, we reproduced four representative deep-learning methods~\cite{miao2020continuous,shaikh2024dual,jamil2024blood,tian2025paralleled}, covering both single-modal and multimodal settings and encompassing CNN-, RNN-, and Transformer-based architectures. For a fair comparison, all methods were trained and evaluated on the same dataset using identical train/validation/test splits, optimizer, batch size, and loss formulation, and their performance was reported under the same evaluation metrics as the proposed model. For PPG-based models, we replaced the original single-channel PPG input with the multi-channel PPG signals available in our dataset while keeping the remaining architectural components unchanged. When implementation details were not fully specified in the original papers (e.g., exact channel widths in intermediate layers or the structure of the regression head), we adopted reasonable configurations that closely follow the authors’ descriptions and performed minor tuning based on validation performance. Whenever not explicitly constrained by the original papers, the initial learning rate, number of training epochs, and other model-specific hyperparameters were chosen to be as consistent as possible with those used for proposed model, with only minor adjustments made when necessary based on validation performance. We emphasize that, due to differences in datasets and subtle implementation details, the absolute numerical results on our dataset are not directly comparable to those reported in the original publications; instead, these reproductions are intended to enable a controlled, head-to-head comparison of existing methods and our model under a unified experimental framework.

\section{Experimental Results and Discussion}
\label{sec:exp}
In this section, we evaluate the proposed six-modal framework on the multi-motion-state dataset and compare it with four representative deep-learning baselines as well as several ablation variants. Unless otherwise specified, all models are trained under the same preprocessing and segmentation pipeline, and the data are split into training, validation, and test sets with a ratio of 7:1:2. The evaluation metrics include the mean absolute error (MAE), standard deviation of error (SDE), and root mean square error (RMSE) for systolic (SBP) and diastolic (DBP) blood pressure.

\subsection{Overall Comparison with State-of-the-art Baseline Methods}
\begin{table*}[t]
	\centering
	\caption{Performance comparison of state-of-the-art baselines and ablation variants for cuffless blood pressure estimation across running, walking, and sitting conditions on the Pulse Transit Time PPG Dataset.}
    \label{tab:bp_results}
	\begin{threeparttable}
		\setlength{\tabcolsep}{3pt}
		\renewcommand{\arraystretch}{1.3}
		
		\begin{tabular}{l C{4.0cm} *{3}{c} *{3}{c}}
			\toprule
			Method & Modalities 
			& \multicolumn{3}{c}{SBP} 
			& \multicolumn{3}{c}{DBP} \\
			\cmidrule(lr){3-5} \cmidrule(lr){6-8}
			& & MAE & SDE & RMSE
			& MAE & SDE & RMSE \\
			\midrule
			\multicolumn{8}{l}{\textit{State-of-the-art baselines}} \\
			\midrule
			
			Miao et al.~\cite{miao2020continuous}
			& ECG & 6.69 & 8.70 & 8.74 & 5.01 & 6.59 & 6.59 \\
			\addlinespace[2.0ex]
			
			Tian et al.~\cite{tian2025paralleled}
			& PPG & 8.52 & 11.23 & 11.26 & 5.84 & 7.38 & 7.41 \\
			
			\multirow{2}{*}{Shaikh \& Forouzanfar~\cite{shaikh2024dual}} 
			& ECG 
			& \multirow{2}{*}{4.98} & \multirow{2}{*}{6.30} & \multirow{2}{*}{6.42}
			& \multirow{2}{*}{4.85} & \multirow{2}{*}{6.27} & \multirow{2}{*}{6.28} \\
			& PPG & & & & & & \\
			\addlinespace[2.0ex]
			
			\multirow{2}{*}{Jamil et al.~\cite{jamil2024blood}}
			& ECG 
			& \multirow{2}{*}{4.50} & \multirow{2}{*}{5.84} & \multirow{2}{*}{5.93}
			& \multirow{2}{*}{4.53} & \multirow{2}{*}{5.94} & \multirow{2}{*}{5.95} \\
			& PPG & & & & & & \\
			\addlinespace[2.0ex]

			\addlinespace[3.0ex]
			\multicolumn{8}{l}{\textit{Ablation variants of proposed method}} \\
			\midrule
			
			{Proposed w/o contrastive loss}
			& PPG & 8.55 & 9.78 & 11.26 & 5.86 & 6.39 & 7.12 \\
			\addlinespace[2.0ex]
			
			{Proposed w/o contrastive loss}
			& ECG & 8.29 & 11.47 & 11.48 & 5.50 & 7.26 & 7.26 \\
			\addlinespace[2.0ex]
			
			\multirow{2}{*}{Proposed w/o contrastive loss}
			& ECG 
			& \multirow{2}{*}{6.79} & \multirow{2}{*}{8.83} & \multirow{2}{*}{8.83}
			& \multirow{2}{*}{5.00} & \multirow{2}{*}{6.32} & \multirow{2}{*}{6.34} \\
			& PPG & & & & & & \\
			\addlinespace[2.0ex]
			
			\multirow{3}{*}{Proposed w/o contrastive loss}
			& ECG 
			& \multirow{3}{*}{5.25} & \multirow{3}{*}{5.90} & \multirow{3}{*}{6.76}
			& \multirow{3}{*}{3.51} & \multirow{3}{*}{4.68} & \multirow{3}{*}{4.69} \\
			& PPG & & & & & & \\
			& PHY & & & & & & \\
			\addlinespace[2.0ex]
			
			\multirow{3}{*}{Proposed w/o MoE head}
			& ECG 
			& \multirow{3}{*}{5.10} & \multirow{3}{*}{6.17} & \multirow{3}{*}{6.60}
			& \multirow{3}{*}{4.21} & \multirow{3}{*}{5.06} & \multirow{3}{*}{5.22} \\
			& PPG & & & & & & \\
			& PHY & & & & & & \\
			\addlinespace[2.0ex]
			
			\multirow{3}{*}{\textbf{Proposed}}
			& ECG 
			& \multirow{3}{*}{$\mathbf{3.60}$} & \multirow{3}{*}{$\mathbf{4.62}$} & \multirow{3}{*}{$\mathbf{4.62}$}
			& \multirow{3}{*}{$\mathbf{3.01}$} & \multirow{3}{*}{$\mathbf{3.93}$} & \multirow{3}{*}{$\mathbf{3.97}$} \\
			& PPG & & & & & & \\
			& PHY & & & & & & \\
			
			\bottomrule
		\end{tabular}
		
		\begin{tablenotes}
			\footnotesize
			\item MAE: mean absolute error; 
			SDE: standard deviation of error; 
			RMSE: root mean square error. 
			All values are reported in mmHg. \\
			PHY denotes the four physical sensor modalities: 
			attachment pressure, sensor temperature, acceleration, and angular velocity.
		\end{tablenotes}
	\end{threeparttable}
\end{table*}

\par Table~\ref{tab:bp_results} reports the performance of four state-of-the-art baseline methods and the proposed model. Among the baselines, the method by Miao \emph{et al.} uses only ECG signals, the method by Tian \emph{et al.} uses only PPG signals, while the methods of Shaikh \& Forouzanfar and Jamil \emph{et al.} exploit both ECG and PPG. As can be seen, the two ECG+PPG baselines clearly outperform the ECG-only and PPG-only models, indicating that fusing multimodal pulse wave information already provides a strong starting point for cuffless BP estimation under our unified experimental setting.
\par Building upon this strong baseline, the proposed model (Proposed, ECG+PPG+PHY) achieves the lowest MAE, SDE, and RMSE for both SBP and DBP among all compared methods. For instance, in SBP estimation, the MAE of the strongest baseline (Jamil \emph{et al.}) is 4.50~mmHg, whereas the proposed method reduces the MAE to 3.60~mmHg. For DBP, the MAE is reduced from 4.53~mmHg to 3.01~mmHg. In relative terms, this corresponds to reductions of approximately 20\% and 34\% in SBP and DBP MAE, respectively, with SDE and RMSE exhibiting similar improvements. These results demonstrate that, under the same data and training protocol, combining six synchronized modalities with semantic alignment and an MoE regression head can further improve the accuracy and robustness of cuffless BP estimation beyond strong existing baselines.

\subsection{Ablation Study: Modality Configuration and Model Components}
\par To assess the contribution of multimodal information and the proposed learning components, we conduct an ablation study on different modality configurations and model variants, as summarized in Table~\ref{tab:bp_results}.

\par We first examine the effect of using different modality combinations when the contrastive loss is removed (w/o contrastive loss). When only PPG or only ECG is used, the performance of our framework is comparable to or slightly worse than that of the corresponding single-modality baselines, with SBP MAE values of 8.55~mmHg and 8.29~mmHg and DBP MAE values around 5--6~mmHg. This behavior can be partly attributed to the stricter parameter and computational budget imposed by the six-modal setting: in order to keep the overall architecture computationally efficient when all six modalities are enabled, each single-branch network is implemented as a relatively compact backbone, rather than matching the full capacity of the standalone single-modality architectures used in the strong baselines. The slight performance gap in the single-modality case is therefore not surprising and does not, by itself, indicate that the underlying branch architectures are intrinsically inferior.

\par As we move from PPG-only to ECG+PPG, both SBP and DBP errors are clearly reduced. Further adding the four physical modalities (attachment pressure, temperature, acceleration, and angular velocity) to form the ECG+PPG+PHY configuration yields additional gains: the SBP MAE decreases from 6.79~mmHg to 5.25~mmHg, while the DBP MAE is reduced from 5.00~mmHg to 3.51~mmHg. Overall, the error metrics exhibit an almost monotonic decrease as more modalities are incorporated. These observations indicate that, in a challenging multi-motion-state scenario involving running, walking, and sitting, additional physical sensor signals can effectively compensate for motion artefacts and posture changes present in ECG and PPG, enabling the model to more robustly capture BP-related physiological patterns. In other words, the main advantage of the proposed framework lies in exploiting the complementarity among ECG, PPG, and multiple physical modalities, rather than merely stacking more complex architectures on top of a single modality.

\par We then analyze the contributions of semantic alignment and the MoE regression head. Table~\ref{tab:bp_results} includes two key variants under the same ECG+PPG+PHY configuration: (i) w/o contrastive loss, which keeps the MoE head but removes the contrastive loss, and (ii) w/o MoE head, which replaces the MoE head with a standard regression head while retaining the contrastive loss. Comparing w/o contrastive loss with the full model shows that adding the contrastive loss on top of the MoE-based architecture leads to marked performance gains: the SBP MAE decreases from 5.25~mmHg to 3.60~mmHg, and the DBP MAE decreases from 3.51~mmHg to 3.01~mmHg, with concomitant reductions in SDE and RMSE. This suggests that cross-modal semantic alignment helps bring ECG, PPG, and physical modalities into a shared representation space, enabling the model to more effectively capture BP-related information that is consistently expressed across different sensors.

\par Conversely, when the MoE head is replaced by a standard regression head (w/o MoE head) while keeping the contrastive loss, both SBP and DBP errors increase compared to the full model. Specifically, the SBP MAE rises from 3.60~mmHg to 5.10~mmHg, and the DBP MAE increases from 3.01~mmHg to 4.21~mmHg. These results indicate that the MoE decoder can adaptively allocate different experts to different samples and motion conditions, thereby enhancing the model's ability to fit and generalize across heterogeneous motion states. Taken together, the ablation study suggests that the observed performance gains are largely associated with two factors: (i) leveraging complementary information across ECG, PPG, and multiple physical modalities while keeping the overall architecture computationally efficient, and (ii) enforcing finer-grained semantic alignment and sample-adaptive modeling via contrastive learning and the MoE head.

\subsection{Visualization Analysis}
\begin{figure*}[htbp]
    \centering
    \subfloat[SBP regression plot]{%
        \includegraphics[width=0.45\textwidth]{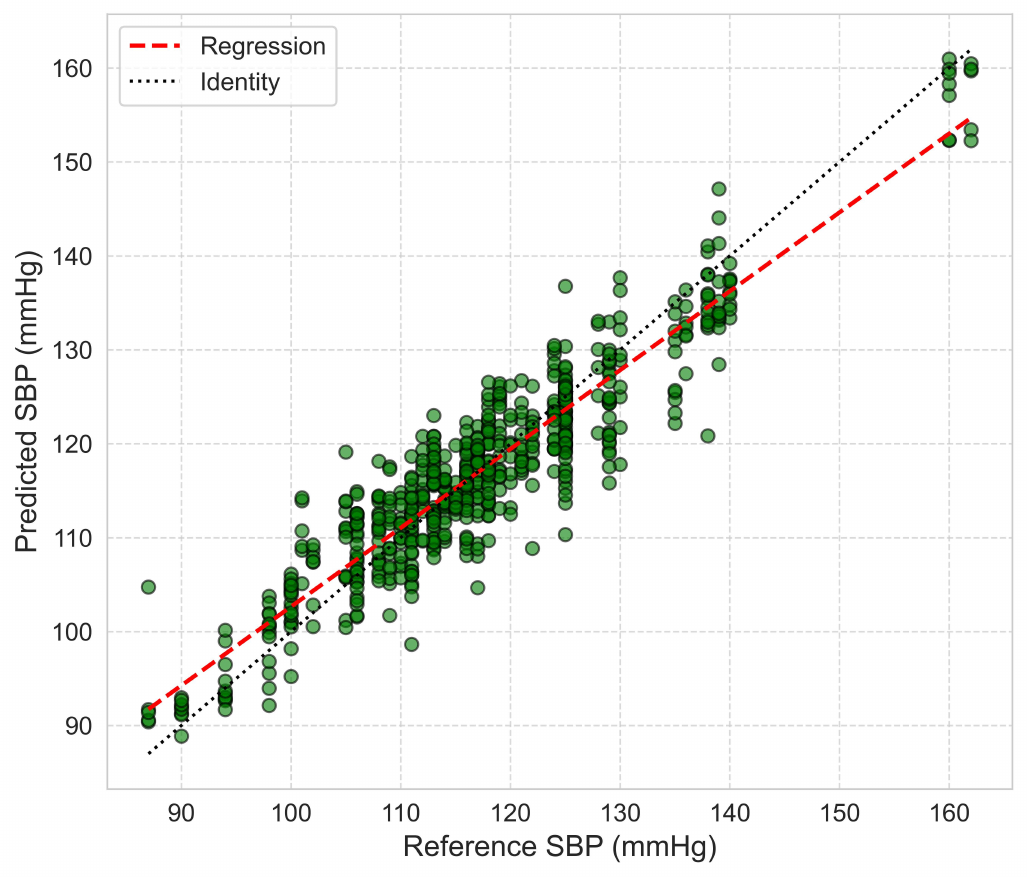}%
        \label{fig:sbp_regression}
    }\hfill
    \subfloat[DBP regression plot]{%
        \includegraphics[width=0.45\textwidth]{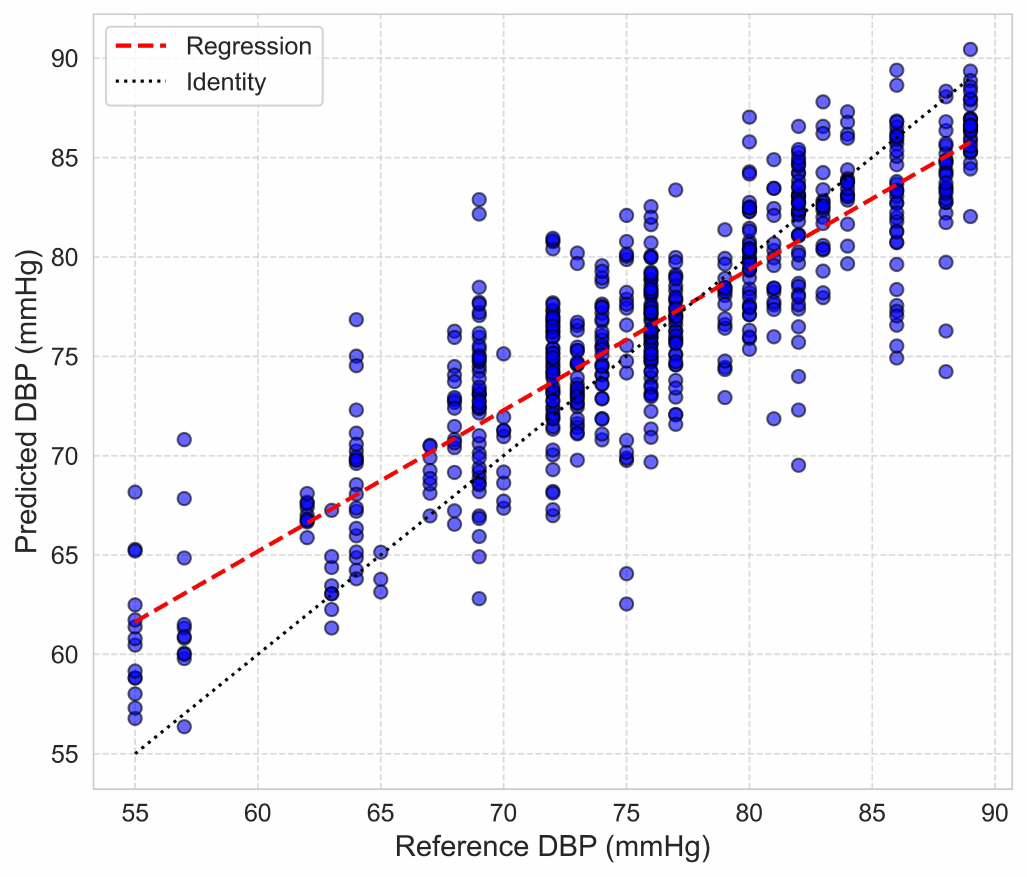}%
        \label{fig:dbp_regression}
    }
    \caption{Regression plots comparing reference and predicted blood pressure values on the test set: (a) systolic blood pressure (SBP) and (b) diastolic blood pressure (DBP). The solid line denotes the identity line, and the dashed line shows the fitted regression.}
    \label{fig:regression_plots}
\end{figure*}

\begin{figure*}[htbp]
    \centering
    \subfloat[SBP Bland--Altman plot]{%
        \includegraphics[width=0.45\textwidth]{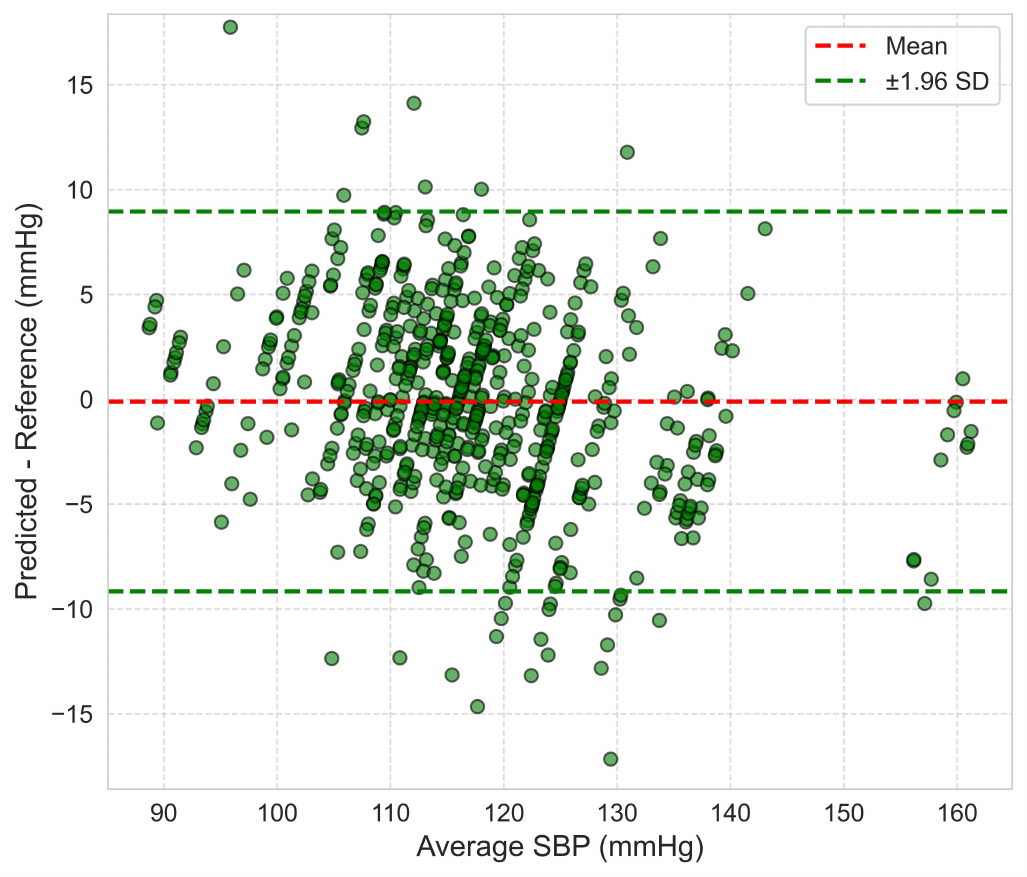}%
        \label{fig:sbp_blandaltman}
    }\hfill
    \subfloat[DBP Bland--Altman plot]{%
        \includegraphics[width=0.45\textwidth]{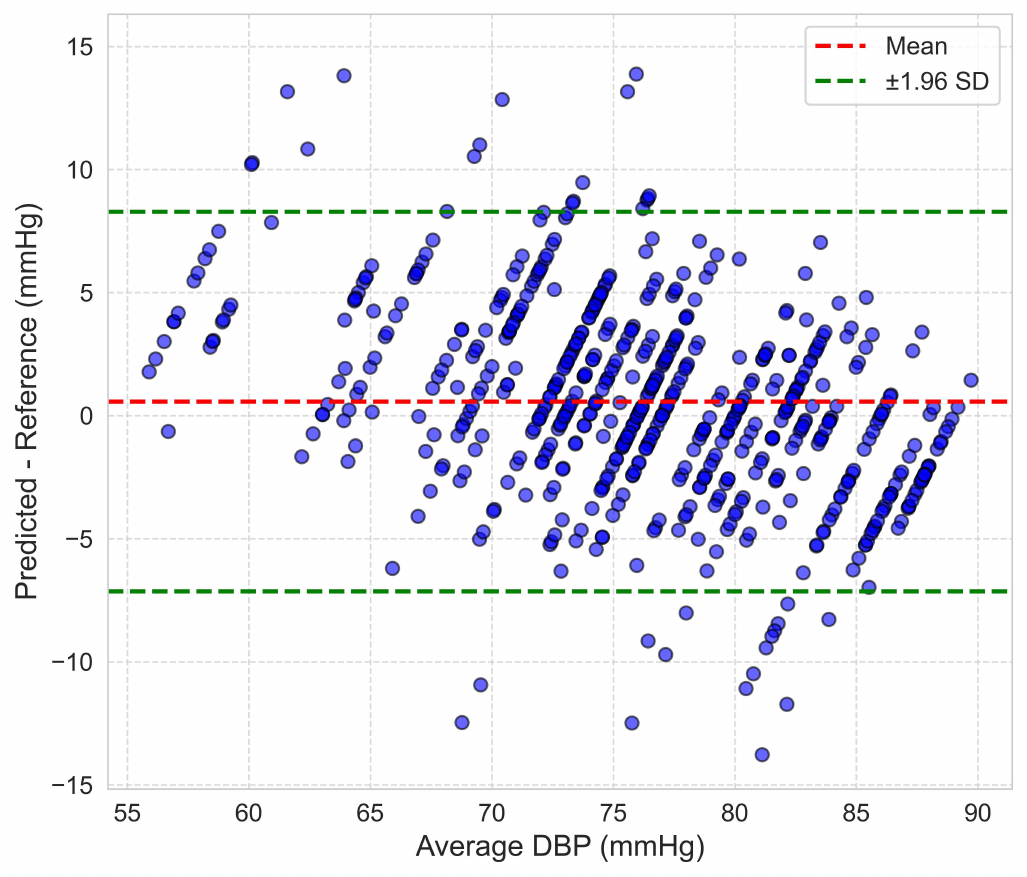}%
        \label{fig:dbp_blandaltman}
    }
    \caption{Bland--Altman plots for the proposed method on the test set: (a) SBP and (b) DBP, showing the mean bias and the 95\% limits of agreement between reference and predicted blood pressure values.}
    \label{fig:blandaltman_plots}
\end{figure*}

\begin{figure*}[t]
    \centering
    \subfloat[SBP error distribution]{%
        \includegraphics[width=0.45\textwidth]{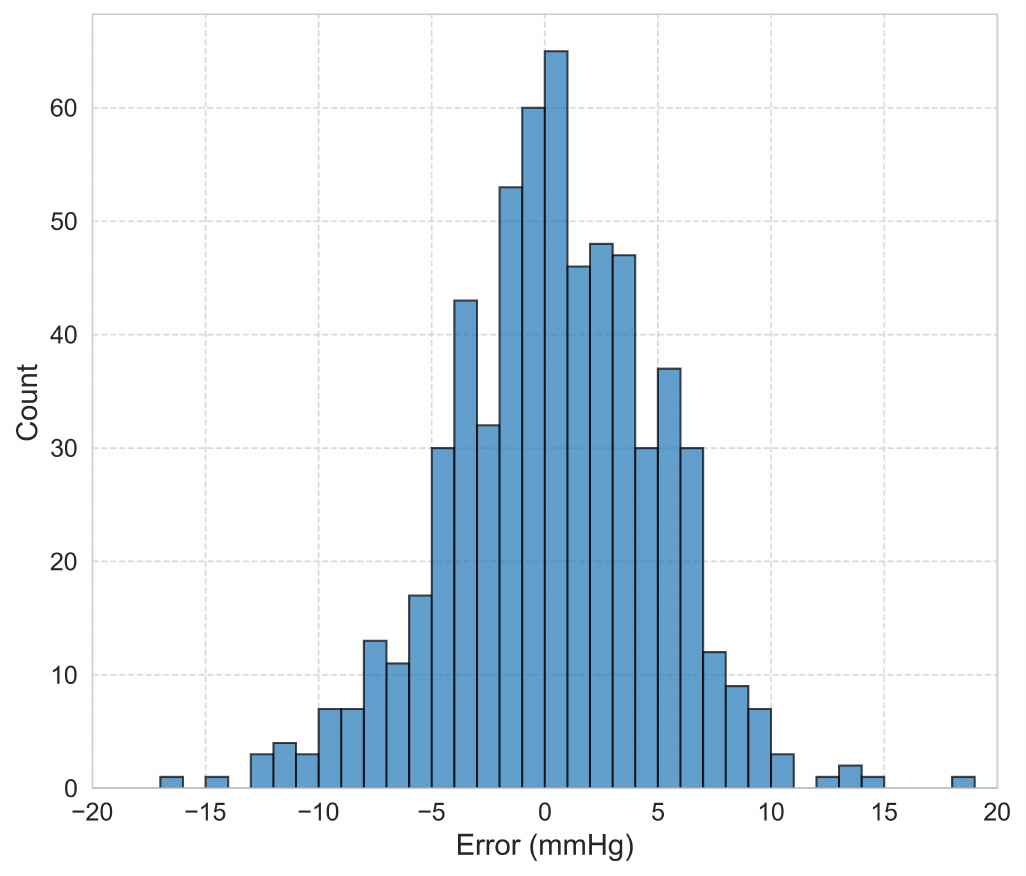}%
        \label{fig:sbp_error_hist}
    }\hfill
    \subfloat[DBP error distribution]{%
        \includegraphics[width=0.45\textwidth]{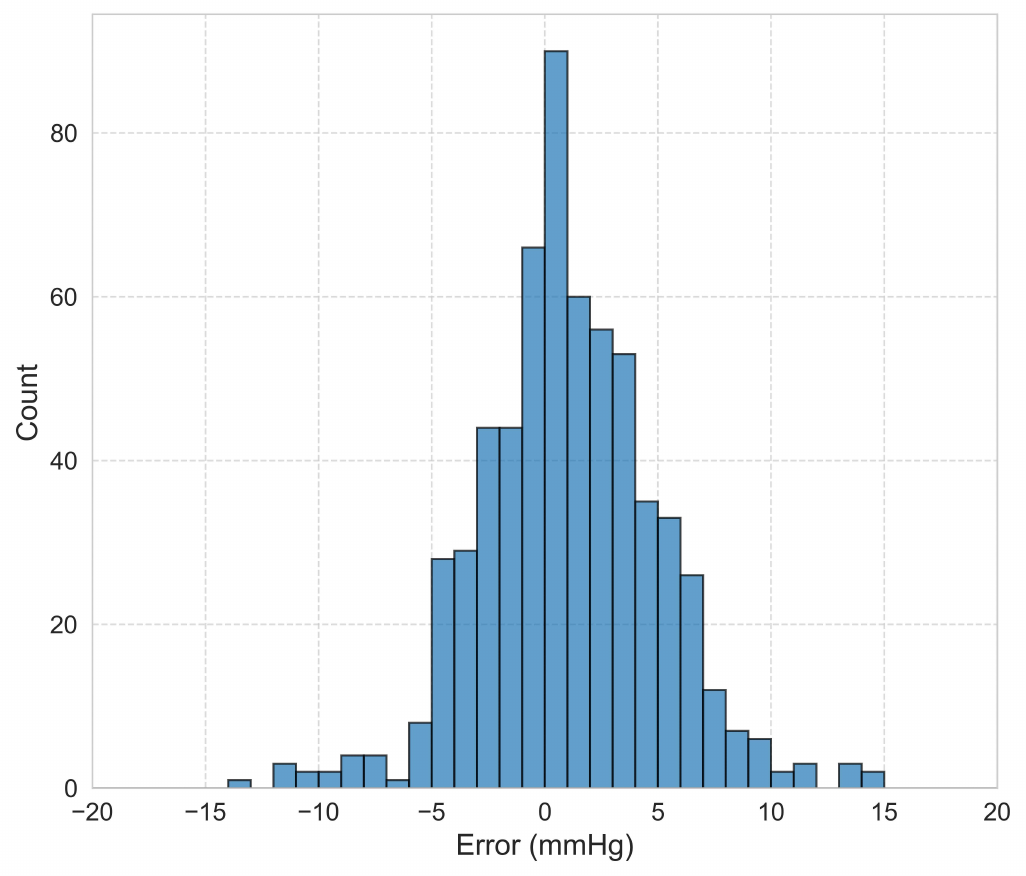}%
        \label{fig:dbp_error_hist}
    }
    \caption{Histograms of estimation errors for (a) systolic blood pressure (SBP) and (b) diastolic blood pressure (DBP) on the test set, illustrating the distribution of residuals produced by the proposed model.}
    \label{fig:error_hist_plots}
\end{figure*}

\par Figure~\ref{fig:regression_plots} shows the regression plots of reference versus predicted SBP and DBP. In both cases, the scatter points are densely concentrated around a regression line that closely follows the identity line, with no obvious saturation or strong nonlinear patterns. This indicates a strong linear association and only a small systematic deviation between predicted and reference BP, which is consistent with the low MAE and RMSE values reported in Table~\ref{tab:bp_results}.

\par Figure~\ref{fig:blandaltman_plots} further presents the Bland--Altman plots. Most points lie within the 95\% limits of agreement, and the mean bias is close to zero for both SBP and DBP. Moreover, no clear trend of increasing bias is observed across low- or high-pressure regions. These observations suggest that the proposed framework does not exhibit substantial systematic overestimation or underestimation over the studied BP range, and that the residual errors are largely dominated by random variability.

\par The error histograms in Figure~\ref{fig:error_hist_plots} provide an additional view of the error distribution. For both SBP and DBP, the empirical error distributions are approximately unimodal and symmetric around zero, with the DBP distribution being noticeably more concentrated, indicating more stable DBP estimation on this dataset. 

\par Taken together, the regression plots, Bland--Altman plots, and error histograms offer an intuitive visual confirmation that the model achieves small and tightly distributed estimation errors for the majority of samples, in line with the reliability suggested by the numerical metrics.

\subsection{Clinical Evaluation: BHS and AAMI Standards}
\begin{table}[htbp]
	\centering
	\caption{British Hypertension Society (BHS) grading criteria for blood pressure measurement devices based on cumulative error percentages.}
    \label{tab:bhs_criteria}
	\begin{tabular}{lccc}
		\toprule
		\multirow{2}{*}{Grade} & \multicolumn{3}{c}{Cumulative error percentage} \\
		\cmidrule(lr){2-4}
		& $\leq 5$ mmHg & $\leq 10$ mmHg & $\leq 15$ mmHg \\
		\midrule
		A & 60\% & 85\% & 95\% \\
		B & 50\% & 75\% & 90\% \\
		C & 40\% & 65\% & 85\% \\
		\bottomrule
	\end{tabular}
\end{table}

\begin{table}[htbp]
	\centering
	\caption{Evaluation of the proposed method according to the BHS grading protocol for systolic, diastolic, and mean arterial pressure.}
    \label{tab:bhs_ctam}
	\begin{tabular}{lcccc}
		\toprule
		\multirow{2}{*}{Parameter} & \multicolumn{3}{c}{Cumulative error percentage} & \multirow{2}{*}{Grade} \\
		\cmidrule(lr){2-4}
		& $\leq 5$ mmHg & $\leq 10$ mmHg & $\leq 15$ mmHg & \\
		\midrule
        SBP & 73.56\% & 96.47\% & 99.68\%  & A \\
		DBP & 82.37\% & 97.28\% & 100.00\% & A \\
		MAP & 85.90\% & 98.40\% & 99.84\%  & A \\
		\bottomrule
	\end{tabular}
\end{table}

\begin{table}[htbp]
	\centering
	\caption{Comparison between the Association for the Advancement of Medical Instrumentation (AAMI) standard requirements and the performance of the proposed method on the Pulse Transit Time PPG Dataset.}
    \label{tab:aami_ctam}
	\begin{threeparttable}
		\setlength{\tabcolsep}{3pt}
		\renewcommand{\arraystretch}{1.2}
		\begin{tabular}{lcccccc}
			\toprule
			\multirow{2}{*}{Parameter} 
			& \multicolumn{2}{c}{ME (mmHg)} 
			& \multicolumn{2}{c}{SD (mmHg)} 
			& \multicolumn{2}{c}{Subjects} \\
			\cmidrule(lr){2-3}\cmidrule(lr){4-5}\cmidrule(lr){6-7}
			& Standard & Ours & Standard & Ours & Standard & Ours \\
			\midrule
			SBP & $\leq 5$ & $-0.11$ & $\leq 8$ & $4.62$ & $\geq 85$ & $22$ \\
			DBP & $\leq 5$ & $0.57$  & $\leq 8$ & $3.93$ & $\geq 85$ & $22$ \\
			\bottomrule
		\end{tabular}
		\begin{tablenotes}
			\footnotesize
			\item ME: mean error; SD: standard deviation of error. 
			The AAMI standard additionally requires evaluation on at least 85 subjects;
			in this study, 22 subjects were available.
		\end{tablenotes}
	\end{threeparttable}
\end{table}

\par To assess the model performance from a clinical perspective, we further examine the results with respect to the BHS and AAMI recommendations.

\par Table~\ref{tab:bhs_criteria} summarizes the cumulative error percentage thresholds for Grades~A, B, and C in the BHS protocol, where Grade~A requires at least 60\%, 85\%, and 95\% of the errors to be within 5, 10, and 15~mmHg, respectively. Table~\ref{tab:bhs_ctam} reports the evaluation of the proposed method. For SBP, DBP, and mean arterial pressure (MAP), the cumulative percentages of errors within 5~mmHg are 73.56\%, 82.37\%, and 85.90\%, respectively; within 10~mmHg, they reach 96.47\%, 97.28\%, and 98.40\%; and within 15~mmHg they approach or reach 100\%. All these values exceed the BHS Grade~A thresholds, indicating that, on this dataset, the method attains Grade~A performance according to the BHS protocol for SBP, DBP, and MAP.

\par Table~\ref{tab:aami_ctam} presents the evaluation with respect to the AAMI standard. AAMI specifies that the absolute mean error (ME) of SBP and DBP should be no greater than 5~mmHg and the standard deviation (SD) of the error should be no greater than 8~mmHg, with validation carried out on at least 85 subjects. In our experiments, the ME and SD for SBP are $-0.11$~mmHg and 4.62~mmHg, respectively, while those for DBP are 0.57~mmHg and 3.93~mmHg, which fall within the numerical ME and SD thresholds specified by AAMI. However, the dataset used in this study contains 22 subjects, which does not meet the minimum sample size requirement of 85 subjects in the AAMI protocol. Therefore, we conclude that the method meets the \emph{numerical} AAMI criteria for ME and SD on this dataset, but the evaluation does not constitute a full AAMI validation because the required number of subjects is not satisfied.

\subsection{Limitations and Future Work}
\par The results above show that, on a multi-motion-state multimodal wearable dataset, the proposed six-modal framework consistently outperforms several strong baselines in terms of MAE, SDE, and RMSE for both SBP and DBP. From a clinical perspective, it attains Grade~A performance under the BHS protocol for SBP, DBP, and MAP, and its mean error and standard deviation fall within the numerical criteria specified by the AAMI standard on this dataset. These findings suggest that, by jointly exploiting ECG, PPG, attachment pressure, temperature, and triaxial acceleration and angular velocity within a unified deep-learning framework with semantic alignment and an MoE regression head, it is possible to achieve cuffless BP estimation with encouraging clinical potential in complex motion scenarios.

\par On the other hand, several limitations remain. First, the Pulse Transit Time PPG Dataset contains 22 healthy subjects, which, although covering running, walking, and sitting states, is still limited in overall scale and may not fully represent broader population variability. Second, because the dataset provides only two cuff-based BP measurements (at the start and end of each recording), all segments extracted from a given subject--state recording are assigned the corresponding end-of-recording cuff values as approximate ground-truth labels for SBP and DBP. While this labeling strategy is a practical approximation under the current data constraints, it may not capture slow intra-recording BP drifts, potentially limiting the model's ability to learn fine-grained BP dynamics. Third, given the modest cohort size and the limited number of reference BP readings per subject, we focus in this work on a unified evaluation setting rather than a strictly subject-independent protocol; as a result, the generalization of the framework to completely unseen individuals and more heterogeneous clinical populations still needs to be confirmed on larger, independent cohorts.

\par In future work, we plan to collect larger-scale, multicenter multimodal wearable datasets with high time-resolution BP labels across more diverse populations and disease conditions, in order to further assess robustness, improve calibration strategies, and validate the generalization and clinical applicability of the proposed framework.

\section{Conclusion}
\label{sec:con}
\par In this paper, we presented a six-modal framework for cuffless blood pressure estimation in multi-motion-state scenarios. The framework integrates ECG, multi-channel PPG, attachment pressure, sensor temperature, and triaxial acceleration and angular velocity on a common temporal axis, using lightweight modality-specific encoders, a contrastive-learning-based semantic alignment module, and a Mixture-of-Experts regression head. Experiments on the Pulse Transit Time PPG Dataset show that the proposed method consistently outperforms several strong deep-learning baselines in terms of MAE, SDE, and RMSE for both SBP and DBP under running, walking, and sitting conditions. From a clinical perspective, it attains Grade~A performance according to the BHS protocol for SBP, DBP, and MAP, and its mean error and standard deviation satisfy the numerical criteria of the AAMI standard on this dataset.

\par This study has several limitations. The dataset contains 22 healthy subjects and only two cuff-based BP measurements per recording, so all segments from a given subject--state share the same approximate BP labels, which constrains both population diversity and the ability to capture slow intra-recording BP drifts. In addition, the proposed framework is evaluated on a single public dataset, and its behavior under different devices, acquisition protocols, and more heterogeneous clinical cohorts remains to be examined. Future work will therefore focus on collecting larger and more diverse multimodal wearable datasets with higher time-resolution BP annotations and on exploring uncertainty estimation, personalized modeling, domain adaptation, and model compression to move closer to robust real-time deployment on resource-constrained wearable devices.

\section*{Acknowledgment}

\par The authors gratefully acknowledge the contributors of the open-access Pulse Transit Time PPG Dataset, which enabled the experimental evaluation in this work.

\bibliographystyle{IEEEtran} 
\bibliography{main.bib}

\end{document}